\documentclass{article}

\PassOptionsToPackage{numbers,sort&compress}{natbib}

\usepackage[final]{neurips_2022}

\usepackage[utf8]{inputenc} %
\usepackage[T1]{fontenc}    %
\usepackage[colorlinks,linkcolor=black,citecolor=blue,urlcolor=black,bookmarks=true]{hyperref}
\usepackage{url}            %
\usepackage{booktabs}       %
\usepackage{amsfonts}       %
\usepackage{nicefrac}       %
\usepackage{microtype}      %
\usepackage{xcolor}         %

\usepackage{amsmath,amsfonts,bm}

\def\eqref#1{equation~\ref{#1}}

\def\1{\bm{1}}

\def\rvk{{\mathbf{k}}}

\def\rvn{{\mathbf{n}}}

\def\rvx{{\mathbf{x}}}

\def\rvz{{\mathbf{z}}}

\def\vtheta{{\bm{\theta}}}
\def\vphi{{\bm{\phi}}}

\def\mI{{\bm{I}}}

\DeclareMathAlphabet{\mathsfit}{\encodingdefault}{\sfdefault}{m}{sl}
\SetMathAlphabet{\mathsfit}{bold}{\encodingdefault}{\sfdefault}{bx}{n}

\def\gN{{\mathcal{N}}}

\newcommand{\pdata}{p_{\rm{data}}}

\newcommand{\E}{\mathbb{E}}

\newcommand{\R}{\mathbb{R}}

\DeclareMathOperator*{\argmax}{arg\,max}
\DeclareMathOperator*{\argmin}{arg\,min}

\usepackage{graphicx}
\usepackage{subcaption}
\usepackage{wrapfig}
\usepackage{orcidlink}
\usepackage{cleveref}
\Crefname{figure}{Fig.}{Figs.}
\Crefname{table}{Tab.}{Tabs.}
\Crefname{section}{Sec.}{Secs.}
\Crefname{appendix}{App.}{Apps.}
\Crefname{equation}{Eq.}{Eqs.}
\Crefname{algorithm}{Alg.}{Algs.}

\newcommand{\TIM}[1]{}

\definecolor{myblue}{HTML}{5983cc}

\title{Latent Space Diffusion Models of Cryo-EM Structures}

\author{Karsten Kreis\textsuperscript{1,}\thanks{Equal contribution.}
\qquad\quad Tim Dockhorn\textsuperscript{1,2,3,*}
\qquad\quad Zihao Li\textsuperscript{4}
\qquad\quad Ellen Zhong\textsuperscript{4}\orcidlink{0000-0001-6345-1907}
\\
\\
{\textsuperscript{1}NVIDIA \quad \textsuperscript{2}University of Waterloo \quad \textsuperscript{3}Vector Institute \quad \textsuperscript{4}Princeton University \vspace{2pt}}
\\
\texttt{\scriptsize kkreis@nvidia.com,} \quad \texttt{\scriptsize tim.dockhorn@uwaterloo.ca,}\quad \texttt{\scriptsize   \{zl1665,zhonge\}@princeton.edu}
}

\begin{document}

\maketitle

\begin{abstract}

Cryo-electron microscopy (cryo-EM) is unique among tools in structural biology in its ability to image large, dynamic protein complexes. 
Key to this ability is image processing algorithms for heterogeneous cryo-EM reconstruction, including recent deep learning-based approaches.
The state-of-the-art method cryoDRGN uses a Variational Autoencoder (VAE) framework to learn a continuous distribution of protein structures from single particle cryo-EM imaging data.
While cryoDRGN can model complex structural motions, the Gaussian prior distribution of the VAE fails to match the aggregate approximate posterior, which prevents generative sampling of structures especially for multi-modal distributions (e.g. compositional heterogeneity).
Here, we train a diffusion model as an expressive, learnable prior in the cryoDRGN framework.
Our approach learns a high-quality generative model over molecular conformations directly from cryo-EM imaging data.
We show the ability to sample from the model on two synthetic and two real datasets, where samples accurately follow the data distribution unlike samples from the VAE prior distribution. 
We also demonstrate how the diffusion model prior can be leveraged for fast latent space traversal and interpolation between states of interest.  
By learning an accurate model of the data distribution, our method unlocks tools in generative modeling, sampling, and distribution analysis for heterogeneous cryo-EM ensembles.

\end{abstract}

\section{Introduction} \label{sec:intro}

Single particle cryo-electron microscopy (cryo-EM) is a biological imaging modality capable of visualizing the 3D structure of large biomolecular complexes at (near-)atomic resolution \cite{Cheng2018-gb}. In this technique, a thin layer of an aqueous sample of the molecule of interest is flash frozen and imaged with a transmission electron microscope. After initial pre-processing of the raw micrographs, the dataset consists of a set of noisy projection images, where each image $\rvx$ contains a 2D projection of a molecular volume $V: \mathbb{R}^3 \rightarrow \mathbb{R}$ captured in an unknown \textit{pose} $(R,t) \in SO(3) \times \mathbb{R}^2$ \cite{Singer2020-rf}. Since each image contains a unique molecule, to account for structural variation between the imaged molecules, reconstruction methods introduce a latent variable $\rvz$ to define a \textit{conformational space} $\mathcal{V}(\cdot,\rvz):\mathbb{R}^3 \rightarrow \mathbb{R}$ from which volumes $V$ are sampled. Thus, \textit{heterogeneous cryo-EM reconstruction} amounts to solving the inverse problem of estimating $\mathcal{V}$, $\rvz$, and $(R,t)$ from the experimental images $\rvx$. Traditionally, common approaches for heterogeneous reconstruction use a discrete model for $\mathcal{V}$ (e.g. a mixture model in 3D classification~\citep{Scheres2010-vi,Lyumkis2013-yi,Punjani2017-ta}); however, deep generative models have recently been introduced that are designed to capture more complex, continuous distributions of protein conformations~\citep{zhong2020Reconstructing, Zhong2021-gn, Zhong2021-nc, Punjani2021-gl, Donnat2022-uj}. \looseness=-1

CryoDRGN~\cite{zhong2020Reconstructing} is a state-of-the-art method for heterogeneous reconstruction based on deep generative modeling. In cryoDRGN, a neural field representation of the volume is conditioned on a generic, continuous latent variable $\rvz \in \mathbb{R}^{\textrm{dim}(\mathbf{z})}$ describing the molecule's conformational space $\mathcal{V}$. CryoDRGN uses a standard Variational Autoencoder (VAE)~\citep{kingma2014vae,rezende2014stochastic} framework for amortized inference of $\rvz$ and reconstruction of $V$. However, the standard Gaussian prior that is employed in regular VAEs to model the latent space distribution cannot accurately match the aggregate posterior distribution (known as the \textit{prior hole problem}~\citep{vahdat2021score,tomczak2018VampPrior,takahashi2019variational,bauer2019resampled,vahdat2020nvae,aneja2020ncpvae,sinha2021d2c,rosca2018distribution,hoffman2016elbo}). The latter can be a complex, multi-modal distribution, in particular for molecules with complex topologies and conformational state spaces. 
Because of this, cryoDRGN is only used as an efficient reconstruction method, but not as a generative model that can synthesize meaningful latents from its prior and generate plausible volumes $V$.

\begin{figure}[t!]
\vspace{-5mm}
\begin{minipage}[c]{0.7\textwidth}
    \centering
    \includegraphics[width=1.0\textwidth]{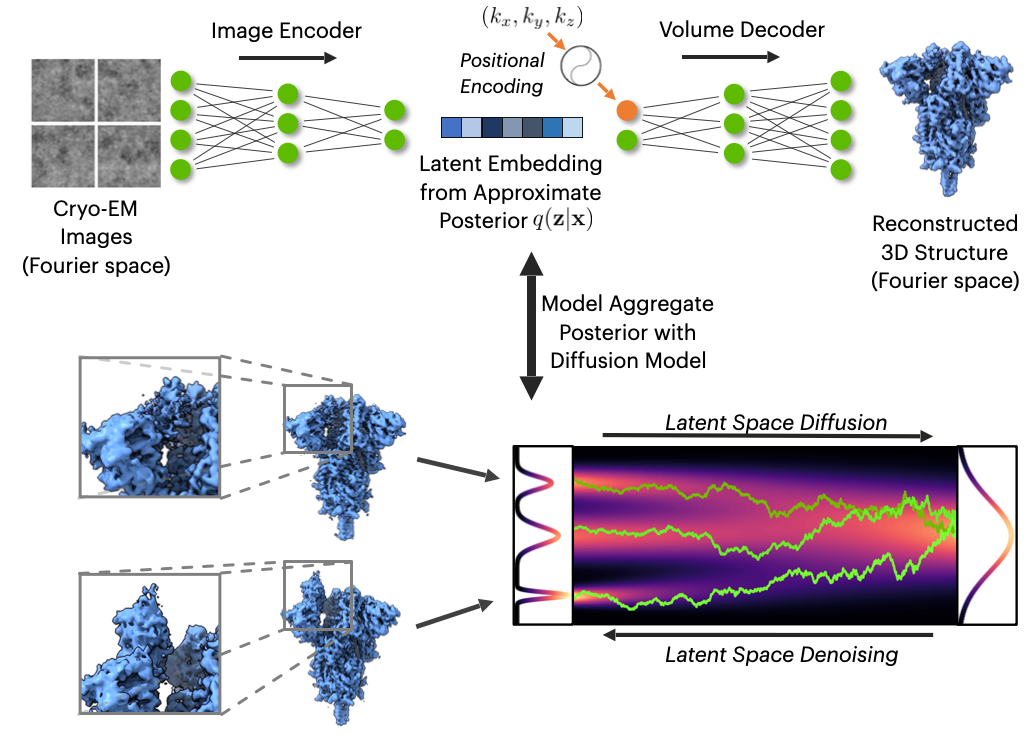}
\end{minipage}\hfill
\begin{minipage}[c]{0.275\textwidth}
    \caption{\small We first train cryoDRGN with a standard VAE objective (\textit{top}). Then, we train a diffusion model in its latent space on the embedding distribution (\textit{bottom}). The diffusion model learns a representation of the conformational space of the molecule. Generating different latents $\rvz$ corresponds to sampling different molecular configurations (e.g., the \textcolor{myblue}{\textbf{blue}} SARS-CoV-2 spike protein is sampled in receptor binding domain open vs. closed states). (Dataset: Walls et al.~\cite{Walls2020-yq}) \looseness=-1}
    \label{fig:pipeline}
\end{minipage}
\vspace{-5mm}
\end{figure}

Recently, \citet{vahdat2021score} leveraged expressive denoising diffusion generative models~\citep{sohl2015,song2020,ho2020} to better capture complex aggregate posteriors in VAEs for RGB image modeling tasks. 
In this work, we propose to leverage a similar latent diffusion model framework in cryoDRGN, and we train a diffusion prior that can effectively encode the conformational states of the molecules (see~\Cref{fig:pipeline}). In experiments on synthetic and real datasets, we verify that our latent diffusion model indeed accurately captures the latent embedding distribution. Moreover, we show how we can leverage the model for accelerated iterative sampling and traversal in latent space as well as for interpolations between different states, thereby effectively simulating conformational transitions. Our approach learns a true generative model of the conformational space directly from cryo-EM imaging data. We envision that our framework paves the way towards several technical extensions and novel applications in protein structure modeling with cryo-EM data.

\section{Methods} \label{sec:method}

\textbf{CryoDRGN}~\citep{zhong2020Reconstructing,Zhong2021-nmeth} is designed as a VAE with an MLP encoder neural network that consumes Fourier-space cryo-EM images $\rvx$ and parametrizes a diagonal Gaussian approximate posterior $q_\vphi(\rvz|\rvx)$ over latent variables $\rvz$. CryoDRGN's probabilistic Gaussian decoder $p_\vphi(\rvx|\rvz)$ is parametrized as a neural field~\cite{Xie2021-cq}, also based on an MLP, taking both latent variables $\rvz$ and Fourier space coordinates $\rvk=(k_x, k_y, k_z)$ (processed through sinusoidal positional encodings~\citep{zhong2020Reconstructing,mildenhall2020nerf}) as inputs. The field outputs the Coulomb scattering potential in Fourier space at $\rvk$. By the image formation model, 2D central slices of the 3D densities in Fourier space correspond to the input cryo-EM images; During training, a 2D grid of oriented $\rvk$ are rendered to form a reconstruction loss.
However, this requires knowledge of the pose of the molecule, as discussed in~\Cref{sec:intro}. Therefore, cryoDRGN also incorporates an image pose inference step~\citep{Zhong2021-gn}. 
Specifically, a global search over rotations and translations is performed to find the maximum likelihood pose under the decoder distribution, given the inferred latent variable $\rvz$. Finally, cryoDRGN is trained with a modified variational lower bound~\citep{kingma2014vae,rezende2014stochastic} (we omit indicating pose inference here for brevity):\looseness=-1
\begin{equation} \label{eq:cryodrgn_obj}
    \argmax\nolimits_{\vphi} \;\E_{\rvx \sim \pdata(\rvx)} \E_{\rvz \sim q_\vphi(\rvz|\rvx)}\left[\log p_\vphi(\rvx|\rvz)\right] - \beta\, \textrm{KL}\left(q_\vphi(\rvz|\rvx)||p(\rvz)\right),
\end{equation}
with standard Gaussian prior $p(\rvz)\sim\gN(\bm{0}, \mI)$. The Kullback Leibler (KL) divergence weighting is chosen to be $\beta=1/\textrm{dim}(\mathbf{z})<1$ in cryoDRGN. This effectively reduces the KL regularization and gives more flexibility to the model to learn diverse encodings that translate to accurate reconstructions (however, it also encourages a mismatch between prior and aggregate approximate posterior; see discussion below).
By sampling the entire neural field via $\rvk$, cryoDRGN can reconstruct 3D cryo-EM density volumes $V$, given a latent variable $\rvz$.\looseness=-1

\textbf{Diffusion Models}~\citep{sohl2015,song2020,ho2020} are a novel class of deep generative models that has recently demonstrated state-of-the-art quality in image synthesis~\citep{nichol21,dhariwal2021diffusion,rombach2021highresolution,saharia2021image,saharia2021palette,pandey2022diffusevae,dockhorn2022scorebased,xiao2022tackling,saharia2022imagen,ramesh2022dalle2} as well as many other applications. In this work, we use continuous-time diffusion models~\citep{song2020} and follow~\citet{karras2022elucidating}. Let $p(\rvx; \sigma)$ denote the distribution obtained by adding i.i.d. $\sigma^2$-variance Gaussian noise to the data distribution $\pdata(\rvx)$.
For sufficiently large $\sigma_\mathrm{max}$,  $p(\rvx; \sigma_\mathrm{max}^2)$ is almost indistinguishable from $\sigma^2_\mathrm{max}$-variance Gaussian noise.
Based on this observation, diffusion models sample high variance Gaussian noise $\rvx_0 \sim\gN\left(\bm{0}, \sigma_\mathrm{max}^2\right)$ and sequentially denoise $\rvx_0$ into $\rvx_i\sim p(\rvx_i; \sigma_i)$, $i \in [0,...,M]$, with $\sigma_{i} < \sigma_{i-1}$ ($\sigma_0=\sigma_\mathrm{max}$). If $\sigma_M=0$, then $\rvx_0$ is distributed according to the data.
In practice, the sequential denoising is often implemented through the numerical simulation of the \emph{Probability Flow} ordinary differential equation (ODE)~\citep{song2020}\looseness=-1
\begin{align} \label{eq:probability_flow_ode}
    d\rvx = -\dot \sigma(t) \sigma(t) \nabla_\rvx \log p(\rvx; \sigma(t)) \, dt,
\end{align}
where $\nabla_\rvx \log p(\rvx; \sigma)$ is the \emph{score function}~\citep{hyvarinen2005scorematching}. The schedule $\sigma(t): [0, 1] \to \R_+$ is user-specified and  $\dot \sigma(t)$ denotes the time derivative of $\sigma(t)$. Alternatively, we may also use a stochastic differential equation (SDE)~\citep{song2020,karras2022elucidating} that 
effectively samples the same distribution (see~\Cref{app:implementation}).
Diffusion model training reduces to learning the score model $s_\vtheta$. The model can, for example, be parameterized as $\nabla_\rvx \log p(\rvx; \sigma) \approx s_\vtheta = (D_\vtheta(\rvx; \sigma) - \rvx)/ \sigma^2$~\citep{karras2022elucidating}, 
where $D_\vtheta$ is a learnable \emph{denoiser} that, given a noisy data point $\rvx + \rvn$, $\rvx \sim \pdata(\rvx)$, $\rvn \sim \gN\left(\bm{0}, \sigma^2\right)$ and conditioned on the noise level $\sigma$, tries to predict the clean 
$\rvx$. The denoiser $D_\vtheta$ can be trained by minimizing an 
$L_2$-loss ($\lambda(\sigma) \colon \R_+ \to \R_+$ is a weighting function)\looseness=-1
\begin{align} \label{eq:diffusion_objective}
    \argmin\nolimits_{\vtheta}\; \E_{\rvx \sim \pdata(\rvx), \sigma \sim p(\sigma), \rvn \sim \gN\left(\bm{0}, \sigma^2\right)} \left[\lambda(\sigma) \|D_\vtheta(\rvx + \rvn, \sigma) - \rvx \|_2^2 \right].
\end{align}
\begin{figure}[t!]
\vspace{-4mm}
    \centering
    \includegraphics[width=\textwidth]{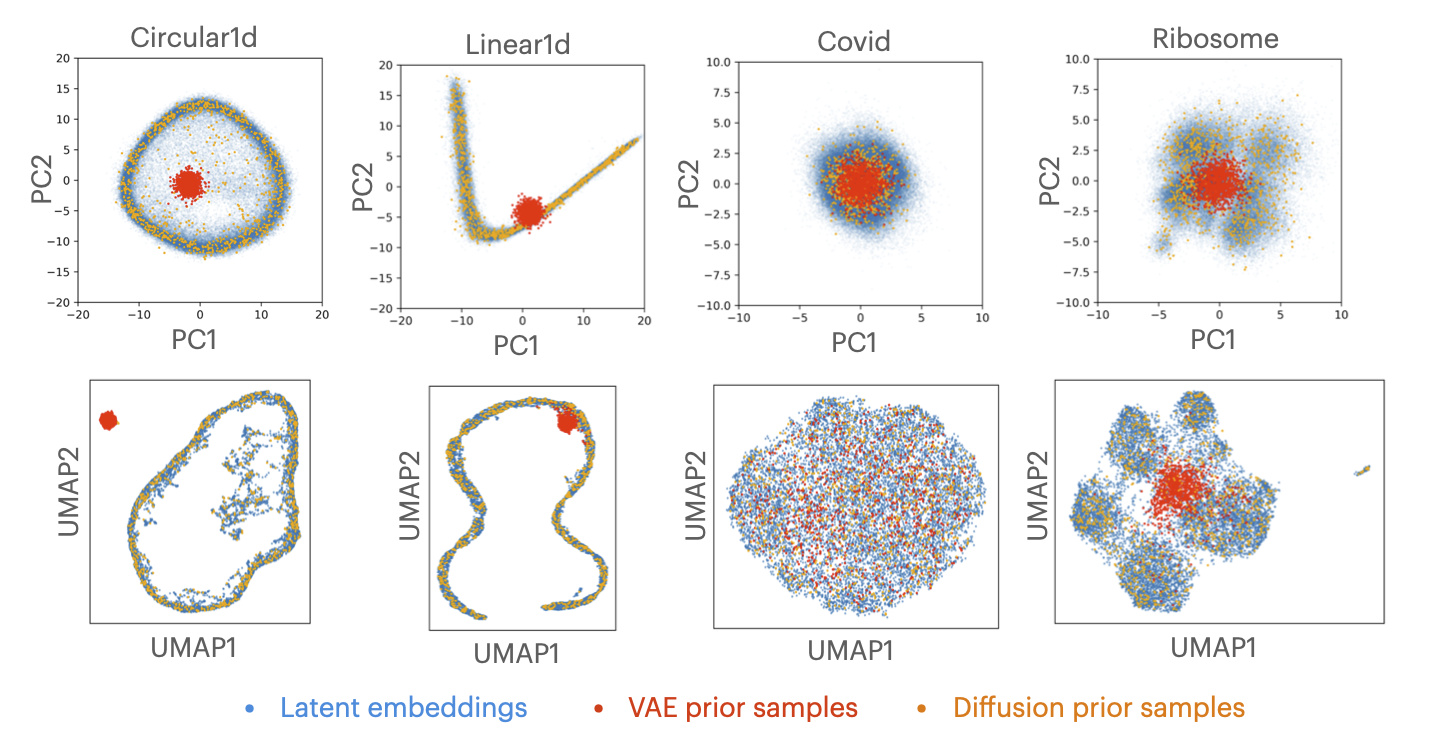}
    \vspace{-1mm}
    \caption{\small Training a diffusion model on cryoDRGN embeddings accurately models the data distribution. CryoDRGN latent embeddings are shown in blue. 1,000 samples from the diffusion model prior or VAE prior are shown in orange or red, respectively. The latent space is visualized in 2D either by projecting samples along the first two principal components of the latent embeddings (\textit{top row}) or with Uniform Manifold Approximation and Projection (UMAP)~\cite{McInnes2018-kt} (\textit{bottom row}). Also see~\Cref{fig:marginals}.}
    \vspace{-3mm}
    \label{fig:prior_samples}
\end{figure}
\textbf{Latent Space Diffusion Models in cryoDRGN.} Since cryoDRGN is trained in regular VAE-fashion with a standard Gaussian prior, it can suffer from the prior hole problem~\cite{vahdat2021score,tomczak2018VampPrior,takahashi2019variational,bauer2019resampled,vahdat2020nvae,aneja2020ncpvae,sinha2021d2c,rosca2018distribution,hoffman2016elbo}, i.e., a mismatch between the aggregate approximate posterior and prior distributions. As we show in~\Cref{sec:results_samples}, samples drawn from the Gaussian prior fail to reproduce the latent embedding distribution. This implies that, although cryoDRGN is a state-of-the-art cryo-EM reconstruction approach, we cannot use it as a generative model and sample latents $\rvz$ according to the modeled distribution of conformational states. Furthermore, prior samples will not necessarily decode into coherent cryo-EM density volumes.

Inspired by recent works on RGB image synthesis that fit diffusion models in the latent space of VAEs~\citep{vahdat2021score,rombach2021highresolution,sinha2021d2c}, we propose to model cryoDRGN's latent embedding distribution with an expressive diffusion model. We can formulate the training objective of our latent space diffusion model as\looseness-1
\begin{align} \label{eq:diffusion_objective2}
    \argmin\nolimits_{\vtheta}\; \E_{\rvx \sim \pdata(\rvx), \rvz \sim q_\vphi(\rvz|\rvx),\sigma \sim p(\sigma), \rvn \sim \gN\left(\bm{0}, \sigma^2\right)} \left[\lambda(\sigma) \|D_\vtheta(\rvz + \rvn, \sigma) - \rvz \|_2^2 \right],
\end{align}
\begin{wraptable}{r}{74mm}
    \caption{\small Total Variation Distance in PC1 marginal space between data embedding distribution and diffusion prior as well as Gaussian VAE prior distributions.}
    \label{tab:tvd}
    \begin{tabular}{@{}lcc@{}}
\toprule
Dataset    & Latent Diffusion & Standard Gaussian \\ 
 & Model Prior & VAE Prior \\
\midrule
\textit{circular1d} & \textbf{0.015}             & 0.82          \\
\textit{linear1d}   & \textbf{0.015}             & 0.86          \\
\textit{covid}      & \textbf{0.005}             & 0.19          \\
\textit{ribosome}   & \textbf{0.015}             & 0.58          \\ \bottomrule
\end{tabular}
\end{wraptable}
where we extract latent embeddings $\rvz$ from the cryo-EM images $\rvx$ with cryoDRGN's encoder $q_\vphi(\rvz|\rvx)$ and fit the diffusion model to the latent space distribution formed by the latent variables $\rvz$. Note that we are training in a two-stage manner: We first train a cryoDRGN VAE model (with the objective in~\Cref{eq:cryodrgn_obj}), which we then freeze. Next, we train the diffusion model in latent space with the objective in~\Cref{eq:diffusion_objective2}. Sampling from the diffusion model and feeding the synthesized latent variables to cryoDRGN's neural field decoder allows us to sample conformational states of the modeled protein. %
Our approach is visualized in~\Cref{fig:pipeline}. All training details are provided in~\Cref{app:implementation}.

\section{Results}
\begin{wrapfigure}{r}{0.45\textwidth}
    \vspace{-16mm}
    \centering
    \includegraphics[width=.45\textwidth]{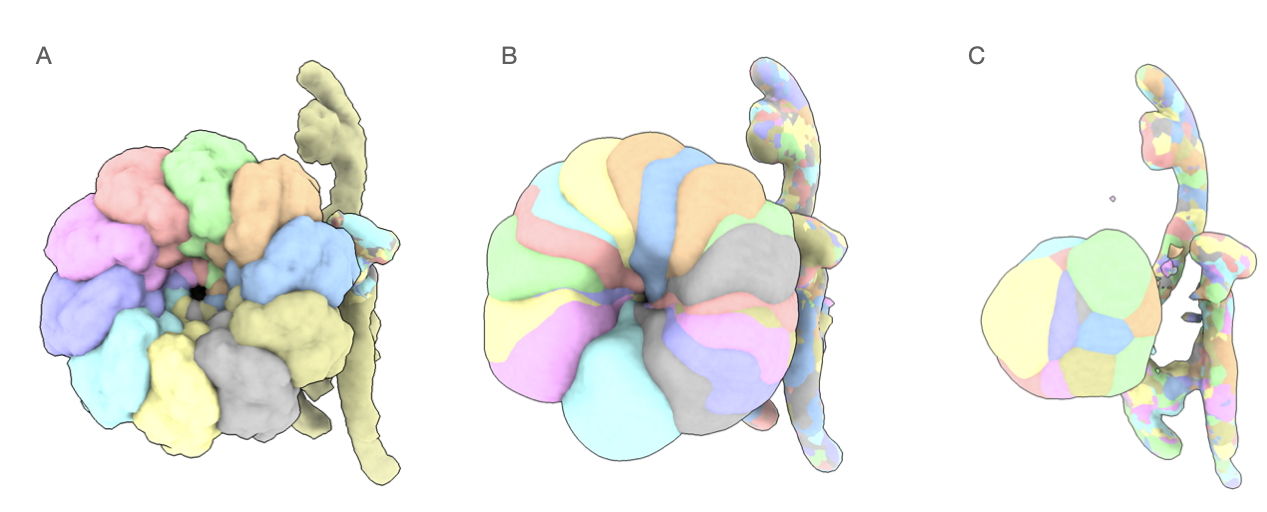}
    \caption{\small \textit{circular1d} volumes viewed as isosurface contour. %
    \textbf{A}: Ground truth volumes. \textbf{B}: 20 volumes sampled from diffusion model prior. \textbf{C}: 20 volumes sampled from standard Gaussian VAE prior.}
    \vspace{-0mm}
    \label{fig:volumes}
\end{wrapfigure}

\subsection{Datasets}
We run experiments on cryoDRGN models trained on a total of four datasets: 2 synthetic datasets, "\textit{linear1d}" and "\textit{circular1d}", where ground truth volumes are generated along a continuous 1-dimensional linear or circular reaction coordinate, respectively~\cite{zhong2020Reconstructing}; "\textit{ribosome}", a real dataset containing a mixture of assembly intermediates of the \textit{E. coli } large ribosomal subunit from EMPIAR-10076~\cite{Davis2016-ug}, and "\textit{covid}", a real dataset of the SARS-CoV-2 spike protein transitioning between the receptor binding domain (RBD) open and closed conformations~\citep{Walls2020-yq}. 
CryoDRGN models were trained with either an 8 or 10 dimensional latent variable model. The sizes of these datasets range from 50k to 277k cryo-EM images. Additional dataset and cryoDRGN training details are provided in~\Cref{app:implementation}. \looseness=-1

\subsection{Generative Model Sampling}
\label{sec:results_samples}
We train a diffusion model on latent encodings from cryoDRGN to model the aggregate approximate posterior distribution for several datasets exhibiting different types of heterogeneity.
\begin{wrapfigure}{l}{0.35\textwidth}
    \vspace{-3mm}
    \centering
    \includegraphics[width=0.28\textwidth]{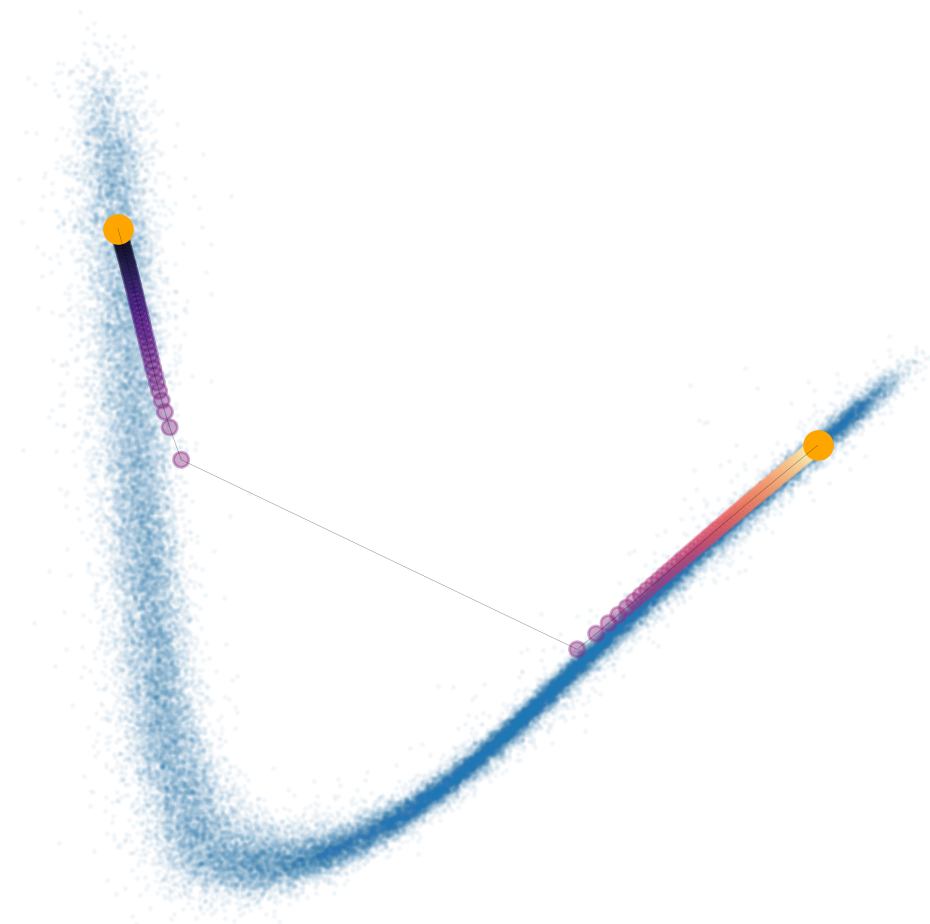}
    \caption{\small Conformational state interpolation with the diffusion model prior for \textit{linear1d}; Also see \Cref{fig:latent_interpolation}.}
    \vspace{-2mm}
    \label{fig:latent_interpolation_linear_main}
\end{wrapfigure}
Once trained, we sample 1,000 latent variables $\rvz$ from the diffusion model (sampling details in~\Cref{app:implementation}). 
A visualization of these samples, together with embeddings of the data distribution and VAE prior samples, is shown in~\Cref{fig:prior_samples}.
While the diffusion model accurately models the data, samples from the original Gaussian VAE prior fail to model the data distribution; the mismatch is worst for highly structured latent spaces.
In \Cref{tab:tvd}, we also show quantitatively that the diffusion model prior captures the distribution of the data's latent encodings much more accurately than the standard Gaussian prior. Furthermore, in~\Cref{fig:volumes} we visualize corresponding 3D volumes for \textit{circular1d}. We find that while the diffusion model samples decode into realistic molecular conformations, the VAE prior samples, which are located in an empty region of the latent space, produce incorrect volumes with artifacts in the heterogeneous region. Our experiments verify that the learned diffusion model prior is able to accurately model the latent embedding distribution.

\subsection{Diffusion Model Interpolations}
We may be interested in modeling transitions between structures by interpolating between their corresponding encodings in latent space. However, as discussed, the latent space distribution can be complex and multi-modal and a linear interpolation directly in latent space between different encodings may pass through prior holes, i.e., ``empty'' regions in latent space, which may produce artifacts. However, our diffusion model effectively maps all states into its own Gaussian prior distribution ($\rvx_0 \sim\gN\left(\bm{0}, \sigma_\mathrm{max}^2\right)$, see~\Cref{sec:method}), and interpolating within this Gaussian space is, in fact, valid. In particular, there are no prior holes in this space, because the diffusion model's forward diffusion process converges to this distribution by construction. Using the deterministic Probability Flow ODE (\Cref{eq:probability_flow_ode}), we encode states $\rvz_A$ and $\rvz_B$ in the diffusion model's own prior (see~\Cref{app:implementation} for details). Denoting the resulting encodings as $\hat{\rvz}_A$ and $\hat{\rvz}_B$, respectively, we then linearly interpolate $\hat{\rvz}_{AB}(y) = (1-y)\hat{\rvz}_A + y\hat{\rvz}_B$, where $y\in[0,1]$ is an interpolation parameter (more sophisticated interpolations, such as spherical, are possible, but we did not observe any benefits). Again using the Probability Flow ODE, we decode states $\hat{\rvz}_{AB}(y)$ along the interpolation back to $\rvz_{AB}(y)$ in cryoDRGN's original latent space. %

\begin{figure}[t!]
\vspace{-5mm}
\begin{minipage}[c]{0.75\textwidth}
    \centering
    \includegraphics[width=1.0\textwidth]{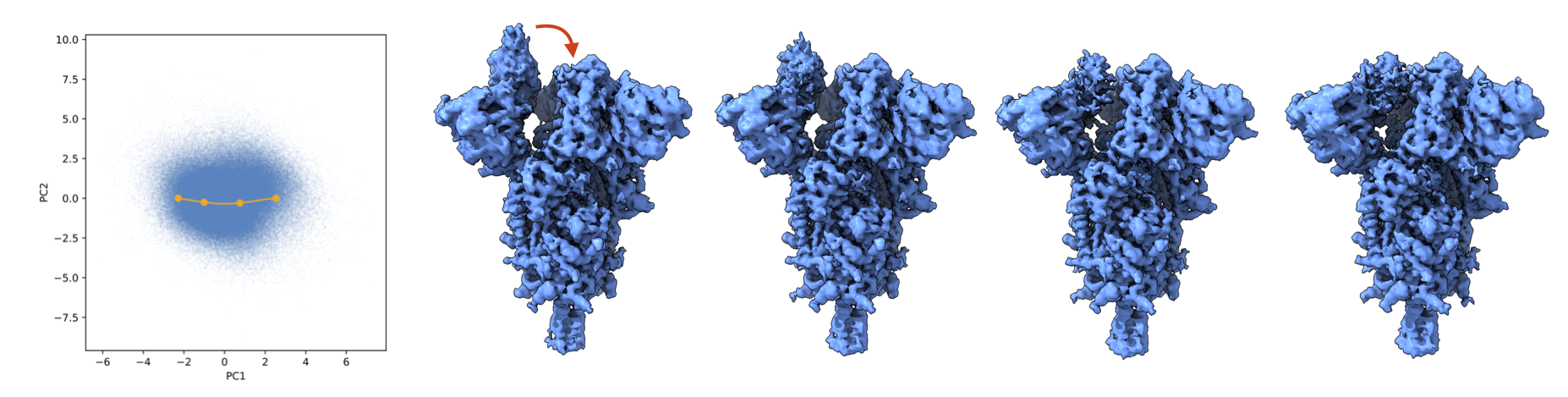}
\end{minipage}\hfill
\begin{minipage}[c]{0.225\textwidth}
    \caption{\small Diffusion model-based interpolation in latent space (\textit{left}) between open and closed conformational states of the SARS-CoV-2 spike protein (visualized on the \textit{right}).}
    \label{fig:covid_interpolation}
\end{minipage}
\vspace{-6mm}
\end{figure}

In \Cref{fig:covid_interpolation}, we visualize such an interpolation trajectory between the open and closed conformations of the SARS-CoV-2 spike protein and we observe a smooth transition. We use cryoDRGN's decoder to generate corresponding molecular volumes. 
We also show the latent space trajectory for an interpolation between two far-apart states in \textit{linear1d}, a challenging case due to its highly structured latent space (\Cref{fig:latent_interpolation_linear_main}). We performed equidistant steps in the diffusion model's Gaussian prior and observe a ``jump'' behavior. This is expected, as it merely implies that the diffusion model has learned a sharp transition in its Probability Flow ODE. Importantly, no samples are generated in the empty region during the jump. 
\looseness=-1

\begin{wrapfigure}{r}{0.35\textwidth}
    \vspace{-18mm}
    \centering
    \includegraphics[width=0.35\textwidth]{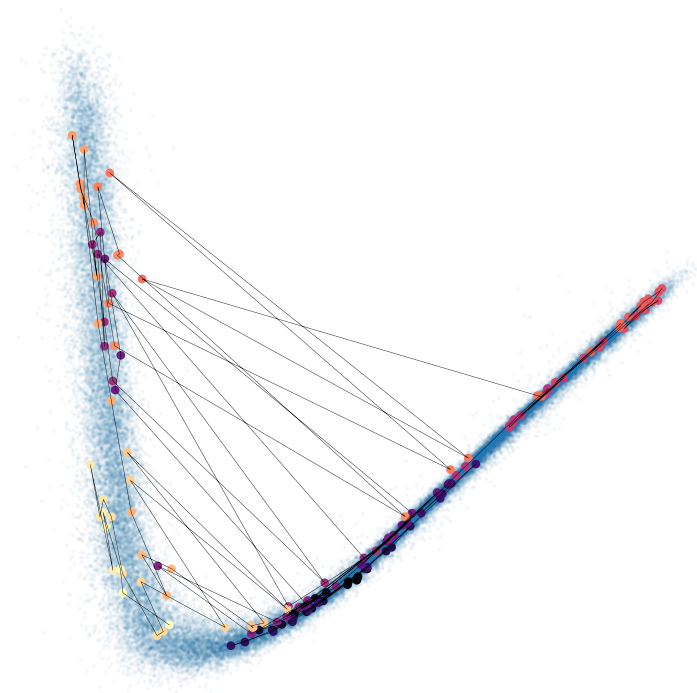}
    \caption{\small Langevin dynamics sampling in the diffusion model prior results in fast exploration of the latent molecular manifold (\textit{linear1d;} also see \Cref{fig:langevin_dynamics}).
    }
    \vspace{-3mm}
    \label{fig:langevin_dynamics_main}
\end{wrapfigure}
\subsection{Efficient Latent Space Traversal and Sampling} \label{sec:langevin}

The diffusion model's Probability Flow ODE can be interpreted as an instance of a continuous Normalizing Flow~\citep{chen2018neuralODE,grathwohl2019ffjord}. Previous work has leveraged Normalizing Flows~\citep{hoffman2019neutra} and other invertible mappings~\citep{xiao2021vaebm} to accelerate Markov Chain Monte Carlo (MCMC) algorithms such as Hamiltonian Monte Carlo and Langevin dynamics~\citep{neal2010hmc}: The MCMC chain is run in the Gaussian prior of the invertible mapping, where no barriers are present, while the Flow non-linearly transforms the samples along the sampling path such that they effectively jump between states and efficiently traverse the complex distribution of interest at the output of the Flow (cryoDRGN's latent space distribution in our case). In \Cref{fig:langevin_dynamics_main}, we demonstrate that running Langevin dynamics in the Gaussian prior space of our latent diffusion model allows us to very efficiently traverse the molecular manifold, making large jumps between distant states. Note that sampling with an iterative sampler like Langevin dynamics can be advantageous, because it would allow us to trivially incorporate additional potential energy terms, which can be more difficult in regular diffusion model sampling.\looseness=-1

\section{Future Directions}
We have shown that a diffusion model incorporated in cryoDRGN's latent space has learned an accurate representation of the conformational state space of the imaged proteins, directly from cryo-EM imaging data. We envision that we can leverage this as a promising tool for relevant and novel applications in generative modeling, accelerated sampling, and distribution analysis in cryo-EM and molecular modeling.

\textbf{Conditional Generative Modeling.} We have demonstrated that random samples from the diffusion model accurately reproduce the molecular manifold. But one may also be able to perform conditional generation: For instance, we envision applications where molecular conformations are sampled from the latent diffusion model based on guidance from auxiliary scoring algorithms or classifiers to fulfill certain properties, like high binding affinity to a ligand. Methodologically, one could build on classifier- and classifier-free guidance techniques from the literature on diffusion models~\citep{song2020,dhariwal2021diffusion,ho2021classifierfree}. Moreover, since we have learned a smooth continuous latent manifold of the data that we can traverse easily, as we demonstrated, we can also employ gradient-based optimization techniques in latent space to optimize the corresponding conformational states according to different criteria~\citep{alperstein2019smiles}.

\textbf{Coupling with Atomic Models.} Future versions of cryoDRGN might perform direct atomic model reconstruction, a direction also explored by~\citet{Zhong2021-nc,rosenbaum2021inferring}. In general, directly inferring an accurate atomic model from cryo-EM imaging data is an extraordinarily challenging problem without any additional inductive biases or priors. However, recently diffusion models have also been used for atomic small molecule and protein generation~\citep{shi2021confgf,xu2022geodiff,jo2022scorebased,wu2022scorebasedmd,jing2022torsional,trippe2022motif,hoogeboom2022equivariant,anand2022protein,wu2022diffusionbased}. Such models can potentially serve as powerful priors for cryo-EM-based atomic protein structure prediction. Combined with an appropriate decoder neural network that operates in atom coordinates, our latent diffusion model would then effectively be a generative model not only over 3D density volumes, but also their underlying atomic structures.

\textbf{Molecular Simulation.} Here, we demonstrated that we can interpolate between different protein states and efficiently traverse the latent space using diffusion model-enhanced MCMC techniques. In fact, MCMC methods like Langevin dynamics are widely used in statistical mechanics and molecular dynamics~\citep{TuckermanBook,leimkuhler2015molecular} to sample molecular conformations. When combined with atomic coordinate outputs, future work could explore leveraging our latent diffusion model for accelerated molecular dynamics simulations as well as for free energy calculations between relevant protein states.

\textbf{End-to-end Training and Hierarchical Models.} There are also promising technical extensions: For instance, end-to-end training of cryoDRGN together with the latent diffusion model, similarly to~\citet{vahdat2021score}, may allow cryoDRGN to learn better embeddings and perform better reconstruction by offering more flexibility to cryoDRGN in how to distribute encodings in latent space. Moreover, deep hierarchical VAEs are often used to improve model expressivity in RGB image synthesis tasks~\citep{vahdat2020nvae,child2021very}. CryoDRGN would likely benefit from similar hierarchical architectures. Such models can sometimes even learn a semantic disentanglement of different features of the data~\citep{preechakul2021diffusion}---exploring this in the context of protein structure modeling would be highly interesting. For instance, we might learn one diffusion model that captures the main data clusters, i.e., main conformational states, and another one to model intra-cluster variation. These directions will be particularly relevant when modeling large-scale, complex and heterogeneous protein data.

In conclusion, we believe that combining state-of-the-art cryo-EM reconstruction methods such as cryoDRGN with modern deep generative learning approaches such as diffusion models has the potential to lead to powerful new tools and promising applications both in cryo-EM reconstruction and in biomolecular modeling more generally.

\bibliographystyle{unsrtnat}
\bibliography{bib_tim}

\begin{thebibliography}{66}
\providecommand{\natexlab}[1]{#1}
\providecommand{\url}[1]{\texttt{#1}}
\expandafter\ifx\csname urlstyle\endcsname\relax
  \providecommand{\doi}[1]{doi: #1}\else
  \providecommand{\doi}{doi: \begingroup \urlstyle{rm}\Url}\fi

\bibitem[Cheng(2018)]{Cheng2018-gb}
Y~Cheng.
\newblock Single-particle {cryo-EM---how} did it get here and where will it go.
\newblock \emph{Science}, 361, 2018.

\bibitem[Singer and Sigworth(2020)]{Singer2020-rf}
Amit Singer and Fred~J Sigworth.
\newblock Computational methods for {Single-Particle} electron cryomicroscopy.
\newblock \emph{Annu Rev Biomed Data Sci}, 3:\penalty0 163--190, July 2020.

\bibitem[Scheres(2010)]{Scheres2010-vi}
Sjors H~W Scheres.
\newblock Chapter eleven - classification of structural heterogeneity by
  {Maximum-Likelihood} methods.
\newblock In Grant~J Jensen, editor, \emph{Methods in Enzymology}, volume 482,
  pages 295--320. Academic Press, January 2010.

\bibitem[Lyumkis et~al.(2013)Lyumkis, Brilot, Theobald, and
  Grigorieff]{Lyumkis2013-yi}
D~Lyumkis, A~F Brilot, D~L Theobald, and N~Grigorieff.
\newblock Likelihood-based classification of {cryo-EM} images using {FREALIGN}.
\newblock \emph{J. Struct. Biol.}, 183, 2013.

\bibitem[Punjani et~al.(2017)Punjani, Rubinstein, Fleet, and
  Brubaker]{Punjani2017-ta}
A~Punjani, J~L Rubinstein, D~J Fleet, and M~Brubaker.
\newblock {cryoSPARC}: algorithms for rapid unsupervised {cryo-EM} structure
  determination.
\newblock \emph{Nat. Methods}, 14, 2017.

\bibitem[Zhong et~al.(2020)Zhong, Bepler, Davis, and
  Berger]{zhong2020Reconstructing}
Ellen~D. Zhong, Tristan Bepler, Joseph~H. Davis, and Bonnie Berger.
\newblock Reconstructing continuous distributions of 3d protein structure from
  cryo-em images.
\newblock In \emph{International Conference on Learning Representations}, 2020.

\bibitem[Zhong et~al.(2021{\natexlab{a}})Zhong, Lerer, Davis, and
  Berger]{Zhong2021-gn}
Ellen~D Zhong, Adam Lerer, Joseph~H Davis, and Bonnie Berger.
\newblock {CryoDRGN2}: Ab initio neural reconstruction of {3D} protein
  structures from real {cryo-EM} images.
\newblock In \emph{Proceedings of the {IEEE/CVF} International Conference on
  Computer Vision}, pages 4066--4075, 2021{\natexlab{a}}.

\bibitem[Zhong et~al.(2021{\natexlab{b}})Zhong, Lerer, Davis, and
  Berger]{Zhong2021-nc}
Ellen~D Zhong, Adam Lerer, Joseph~H Davis, and Bonnie Berger.
\newblock Exploring generative atomic models in {cryo-EM} reconstruction.
\newblock July 2021{\natexlab{b}}.

\bibitem[Punjani and Fleet(2021)]{Punjani2021-gl}
A~Punjani and D~J Fleet.
\newblock {3D} flexible refinement: Structure and motion of flexible proteins
  from {Cryo-EM}.
\newblock \emph{bioRxiv}, 2021.

\bibitem[Donnat et~al.(2022)Donnat, Levy, Poitevin, Zhong, and
  Miolane]{Donnat2022-uj}
Claire Donnat, Axel Levy, Frederic Poitevin, Ellen Zhong, and Nina Miolane.
\newblock Deep generative modeling for volume reconstruction in {Cryo-Electron}
  microscopy.
\newblock January 2022.

\bibitem[Kingma and Welling(2014)]{kingma2014vae}
Diederik~P Kingma and Max Welling.
\newblock Auto-encoding variational bayes.
\newblock In \emph{The International Conference on Learning Representations
  (ICLR)}, 2014.

\bibitem[Rezende et~al.(2014)Rezende, Mohamed, and
  Wierstra]{rezende2014stochastic}
Danilo~Jimenez Rezende, Shakir Mohamed, and Daan Wierstra.
\newblock Stochastic backpropagation and approximate inference in deep
  generative models.
\newblock In \emph{International Conference on Machine Learning}, pages
  1278--1286, 2014.

\bibitem[Vahdat et~al.(2021)Vahdat, Kreis, and Kautz]{vahdat2021score}
Arash Vahdat, Karsten Kreis, and Jan Kautz.
\newblock \href{https://arxiv.org/abs/2106.05931}{Score-based {G}enerative
  {M}odeling in {L}atent {S}pace}.
\newblock In \emph{Neural Information Processing Systems (NeurIPS)}, 2021.

\bibitem[Tomczak and Welling(2018)]{tomczak2018VampPrior}
Jakub Tomczak and Max Welling.
\newblock Vae with a vampprior.
\newblock In \emph{International Conference on Artificial Intelligence and
  Statistics}, pages 1214--1223, 2018.

\bibitem[Takahashi et~al.(2019)Takahashi, Iwata, Yamanaka, Yamada, and
  Yagi]{takahashi2019variational}
Hiroshi Takahashi, Tomoharu Iwata, Yuki Yamanaka, Masanori Yamada, and Satoshi
  Yagi.
\newblock Variational autoencoder with implicit optimal priors.
\newblock \emph{Proceedings of the AAAI Conference on Artificial Intelligence},
  33\penalty0 (01):\penalty0 5066--5073, Jul. 2019.

\bibitem[Bauer and Mnih(2019)]{bauer2019resampled}
Matthias Bauer and Andriy Mnih.
\newblock Resampled priors for variational autoencoders.
\newblock In Kamalika Chaudhuri and Masashi Sugiyama, editors,
  \emph{Proceedings of the Twenty-Second International Conference on Artificial
  Intelligence and Statistics}, volume~89 of \emph{Proceedings of Machine
  Learning Research}, pages 66--75. PMLR, 16--18 Apr 2019.

\bibitem[Vahdat and Kautz(2020)]{vahdat2020nvae}
Arash Vahdat and Jan Kautz.
\newblock \href{https://arxiv.org/abs/2007.03898}{{NVAE: A Deep Hierarchical
  Variational Autoencoder}}.
\newblock In \emph{Neural Information Processing Systems (NeurIPS)}, 2020.

\bibitem[Aneja et~al.(2021)Aneja, Schwing, Kautz, and Vahdat]{aneja2020ncpvae}
Jyoti Aneja, Alexander Schwing, Jan Kautz, and Arash Vahdat.
\newblock {NCP-VAE}: Variational autoencoders with noise contrastive priors.
\newblock In \emph{Advances in Neural Information Processing Systems}, 2021.

\bibitem[Sinha et~al.(2021)Sinha, Song, Meng, and Ermon]{sinha2021d2c}
Abhishek Sinha, Jiaming Song, Chenlin Meng, and Stefano Ermon.
\newblock \href{https://arxiv.org/abs/2106.06819}{{D2C: Diffusion-Denoising
  Models for Few-shot Conditional Generation}}.
\newblock \emph{arXiv:2106.06819}, 2021.

\bibitem[Rosca et~al.(2018)Rosca, Lakshminarayanan, and
  Mohamed]{rosca2018distribution}
Mihaela Rosca, Balaji Lakshminarayanan, and Shakir Mohamed.
\newblock Distribution matching in variational inference.
\newblock \emph{arXiv preprint arXiv:1802.06847}, 2018.

\bibitem[Hoffman and Johnson(2016)]{hoffman2016elbo}
Matthew~D Hoffman and Matthew~J Johnson.
\newblock Elbo surgery: yet another way to carve up the variational evidence
  lower bound.
\newblock In \emph{Workshop in Advances in Approximate Bayesian Inference,
  NeurIPS}, 2016.

\bibitem[Sohl-Dickstein et~al.(2015)Sohl-Dickstein, Weiss, Maheswaranathan, and
  Ganguli]{sohl2015}
Jascha Sohl-Dickstein, Eric Weiss, Niru Maheswaranathan, and Surya Ganguli.
\newblock \href{http://proceedings.mlr.press/v37/sohl-dickstein15.html}{Deep
  {U}nsupervised {L}earning using {N}onequilibrium {T}hermodynamics}.
\newblock In \emph{International Conference on Machine Learning}, 2015.

\bibitem[Song et~al.(2021)Song, Sohl-Dickstein, Kingma, Kumar, Ermon, and
  Poole]{song2020}
Yang Song, Jascha Sohl-Dickstein, Diederik~P Kingma, Abhishek Kumar, Stefano
  Ermon, and Ben Poole.
\newblock \href{https://openreview.net/forum?id=PxTIG12RRHS}{Score-{B}ased
  {G}enerative {M}odeling through {S}tochastic {D}ifferential {E}quations}.
\newblock In \emph{International Conference on Learning Representations}, 2021.

\bibitem[Ho et~al.(2020)Ho, Jain, and Abbeel]{ho2020}
Jonathan Ho, Ajay Jain, and Pieter Abbeel.
\newblock
  \href{https://proceedings.neurips.cc/paper/2020/hash/4c5bcfec8584af0d967f1ab10179ca4b-Abstract.html}{Denoising
  {D}iffusion {P}robabilistic {M}odels}.
\newblock In \emph{Advances in Neural Information Processing Systems}, 2020.

\bibitem[Walls et~al.(2020)Walls, Park, Tortorici, Wall, McGuire, and
  Veesler]{Walls2020-yq}
Alexandra~C Walls, Young-Jun Park, M~Alejandra Tortorici, Abigail Wall,
  Andrew~T McGuire, and David Veesler.
\newblock Structure, function, and antigenicity of the {SARS-CoV-2} spike
  glycoprotein.
\newblock \emph{Cell}, 181\penalty0 (2):\penalty0 281--292.e6, apr 2020.

\bibitem[Zhong et~al.(2021{\natexlab{c}})Zhong, Bepler, Berger, and
  Davis]{Zhong2021-nmeth}
Ellen~D Zhong, Tristan Bepler, Bonnie Berger, and Joseph~H Davis.
\newblock {CryoDRGN}: reconstruction of heterogeneous {cryo-EM} structures
  using neural networks.
\newblock \emph{Nat. Methods}, 18\penalty0 (2):\penalty0 176--185, February
  2021{\natexlab{c}}.

\bibitem[Xie et~al.(2021)Xie, Takikawa, Saito, Litany, Yan, Khan, Tombari,
  Tompkin, Sitzmann, and Sridhar]{Xie2021-cq}
Yiheng Xie, Towaki Takikawa, Shunsuke Saito, Or~Litany, Shiqin Yan, Numair
  Khan, Federico Tombari, James Tompkin, Vincent Sitzmann, and Srinath Sridhar.
\newblock Neural fields in visual computing and beyond.
\newblock November 2021.

\bibitem[Mildenhall et~al.(2020)Mildenhall, Srinivasan, Tancik, Barron,
  Ramamoorthi, and Ng]{mildenhall2020nerf}
Ben Mildenhall, Pratul~P. Srinivasan, Matthew Tancik, Jonathan~T. Barron, Ravi
  Ramamoorthi, and Ren Ng.
\newblock Nerf: Representing scenes as neural radiance fields for view
  synthesis.
\newblock In \emph{ECCV}, 2020.

\bibitem[Nichol and Dhariwal(2021)]{nichol21}
Alexander~Quinn Nichol and Prafulla Dhariwal.
\newblock \href{http://proceedings.mlr.press/v139/nichol21a.html}{Improved
  {D}enoising {D}iffusion {P}robabilistic {M}odels}.
\newblock In \emph{International Conference on Machine Learning}, 2021.

\bibitem[Dhariwal and Nichol(2021)]{dhariwal2021diffusion}
Prafulla Dhariwal and Alex Nichol.
\newblock \href{https://openreview.net/forum?id=OU98jZWS3x_}{Diffusion {M}odels
  {B}eat {GAN}s on {I}mage {S}ynthesis}.
\newblock In \emph{Neural Information Processing Systems}, 2021.

\bibitem[Rombach et~al.(2021)Rombach, Blattmann, Lorenz, Esser, and
  Ommer]{rombach2021highresolution}
Robin Rombach, Andreas Blattmann, Dominik Lorenz, Patrick Esser, and Björn
  Ommer.
\newblock \href{https://arxiv.org/abs/2112.10752}{{High-Resolution Image
  Synthesis with Latent Diffusion Models}}.
\newblock \emph{arXiv:2112.10752}, 2021.

\bibitem[Saharia et~al.(2021{\natexlab{a}})Saharia, Ho, Chan, Salimans, Fleet,
  and Norouzi]{saharia2021image}
Chitwan Saharia, Jonathan Ho, William Chan, Tim Salimans, David~J Fleet, and
  Mohammad Norouzi.
\newblock \href{https://arxiv.org/abs/2104.07636}{Image {S}uper-{R}esolution
  via {I}terative {R}efinement}.
\newblock \emph{arXiv:2104.07636}, 2021{\natexlab{a}}.

\bibitem[Saharia et~al.(2021{\natexlab{b}})Saharia, Chan, Chang, Lee, Ho,
  Salimans, Fleet, and Norouzi]{saharia2021palette}
Chitwan Saharia, William Chan, Huiwen Chang, Chris~A. Lee, Jonathan Ho, Tim
  Salimans, David~J. Fleet, and Mohammad Norouzi.
\newblock \href{https://arxiv.org/abs/2111.05826}{{Palette: Image-to-Image
  Diffusion Models}}.
\newblock \emph{arXiv:2111.05826}, 2021{\natexlab{b}}.

\bibitem[Pandey et~al.(2022)Pandey, Mukherjee, Rai, and
  Kumar]{pandey2022diffusevae}
Kushagra Pandey, Avideep Mukherjee, Piyush Rai, and Abhishek Kumar.
\newblock \href{https://arxiv.org/abs/2201.00308}{{DiffuseVAE: Efficient,
  Controllable and High-Fidelity Generation from Low-Dimensional Latents}}.
\newblock \emph{arXiv:2201.00308}, 2022.

\bibitem[Dockhorn et~al.(2022)Dockhorn, Vahdat, and
  Kreis]{dockhorn2022scorebased}
Tim Dockhorn, Arash Vahdat, and Karsten Kreis.
\newblock \href{https://openreview.net/forum?id=CzceR82CYc}{{Score-Based
  Generative Modeling with Critically-Damped Langevin Diffusion}}.
\newblock In \emph{International Conference on Learning Representations}, 2022.

\bibitem[Xiao et~al.(2022)Xiao, Kreis, and Vahdat]{xiao2022tackling}
Zhisheng Xiao, Karsten Kreis, and Arash Vahdat.
\newblock \href{https://openreview.net/forum?id=JprM0p-q0Co}{{Tackling the
  Generative Learning Trilemma with Denoising Diffusion GANs}}.
\newblock In \emph{International Conference on Learning Representations}, 2022.

\bibitem[Saharia et~al.(2022)Saharia, Chan, Saxena, Li, Whang, Denton,
  Ghasemipour, Ayan, Mahdavi, Lopes, et~al.]{saharia2022imagen}
Chitwan Saharia, William Chan, Saurabh Saxena, Lala Li, Jay Whang, Emily
  Denton, Seyed Kamyar~Seyed Ghasemipour, Burcu~Karagol Ayan, S~Sara Mahdavi,
  Rapha~Gontijo Lopes, et~al.
\newblock \href{https://arxiv.org/abs/2205.11487}{{Photorealistic Text-to-Image
  Diffusion Models with Deep Language Understanding}}.
\newblock \emph{arXiv:2205.11487}, 2022.

\bibitem[Ramesh et~al.(2022)Ramesh, Dhariwal, Nichol, Chu, and
  Chen]{ramesh2022dalle2}
Aditya Ramesh, Prafulla Dhariwal, Alex Nichol, Casey Chu, and Mark Chen.
\newblock \href{https://arxiv.org/abs/2204.06125}{{Hierarchical
  Text-Conditional Image Generation with CLIP Latents}}.
\newblock \emph{arXiv:2204.06125}, 2022.

\bibitem[Karras et~al.(2022)Karras, Aittala, Aila, and
  Laine]{karras2022elucidating}
Tero Karras, Miika Aittala, Timo Aila, and Samuli Laine.
\newblock \href{https://arxiv.org/abs/2206.00364}{{Elucidating the Design Space
  of Diffusion-Based Generative Models}}.
\newblock \emph{arXiv:2206.00364}, 2022.

\bibitem[Hyv\"{a}rinen(2005)]{hyvarinen2005scorematching}
Aapo Hyv\"{a}rinen.
\newblock \href{https://www.jmlr.org/papers/v6/hyvarinen05a.html}{{Estimation
  of Non-Normalized Statistical Models by Score Matching}}.
\newblock \emph{Journal of Machine Learning Research}, 6:\penalty0 695–709,
  2005.
\newblock ISSN 1532-4435.

\bibitem[McInnes et~al.(2018)McInnes, Healy, and Melville]{McInnes2018-kt}
Leland McInnes, John Healy, and James Melville.
\newblock {UMAP}: Uniform manifold approximation and projection for dimension
  reduction.
\newblock February 2018.

\bibitem[Davis(2016)]{Davis2016-ug}
J~H Davis.
\newblock Modular assembly of the bacterial large ribosomal subunit.
\newblock \emph{Cell}, 167, 2016.

\bibitem[Chen et~al.(2018)Chen, Rubanova, Bettencourt, and
  Duvenaud]{chen2018neuralODE}
Ricky T.~Q. Chen, Yulia Rubanova, Jesse Bettencourt, and David Duvenaud.
\newblock
  \href{https://papers.nips.cc/paper/2018/hash/69386f6bb1dfed68692a24c8686939b9-Abstract.html}{{Neural
  Ordinary Differential Equations}}.
\newblock \emph{Advances in Neural Information Processing Systems}, 2018.

\bibitem[Grathwohl et~al.(2019)Grathwohl, Chen, Bettencourt, Sutskever, and
  Duvenaud]{grathwohl2019ffjord}
Will Grathwohl, Ricky T.~Q. Chen, Jesse Bettencourt, Ilya Sutskever, and David
  Duvenaud.
\newblock \href{https://openreview.net/forum?id=rJxgknCcK7}{{FFJORD: Free-form
  Continuous Dynamics for Scalable Reversible Generative Models}}.
\newblock \emph{International Conference on Learning Representations}, 2019.

\bibitem[Hoffman et~al.(2019)Hoffman, Sountsov, Dillon, Langmore, Tran, and
  Vasudevan]{hoffman2019neutra}
Matthew Hoffman, Pavel Sountsov, Joshua~V. Dillon, Ian Langmore, Dustin Tran,
  and Srinivas Vasudevan.
\newblock Neutra-lizing bad geometry in hamiltonian monte carlo using neural
  transport.
\newblock \emph{arXiv:1903.03704}, 2019.

\bibitem[Xiao et~al.(2021)Xiao, Kreis, Kautz, and Vahdat]{xiao2021vaebm}
Zhisheng Xiao, Karsten Kreis, Jan Kautz, and Arash Vahdat.
\newblock \href{https://openreview.net/forum?id=5m3SEczOV8L}{{VAEBM: A
  Symbiosis between Variational Autoencoders and Energy-based Models}}.
\newblock In \emph{International Conference on Learning Representations}, 2021.

\bibitem[Neal(2011)]{neal2010hmc}
Radford~M. Neal.
\newblock {MCMC} {U}sing {H}amiltonian {D}ynamics.
\newblock \emph{Handbook of Markov Chain Monte Carlo}, 54:\penalty0 113--162,
  2011.

\bibitem[Ho and Salimans(2021)]{ho2021classifierfree}
Jonathan Ho and Tim Salimans.
\newblock \href{https://openreview.net/forum?id=qw8AKxfYbI}{{Classifier-Free
  Diffusion Guidance}}.
\newblock In \emph{NeurIPS 2021 Workshop on Deep Generative Models and
  Downstream Applications}, 2021.

\bibitem[Alperstein et~al.(2019)Alperstein, Cherkasov, and
  Rolfe]{alperstein2019smiles}
Zaccary Alperstein, Artem Cherkasov, and Jason~Tyler Rolfe.
\newblock All smiles variational autoencoder.
\newblock \emph{arXiv preprint arXiv:1905.13343}, 2019.

\bibitem[Rosenbaum et~al.(2021)Rosenbaum, Garnelo, Zielinski, Beattie, Clancy,
  Huber, Kohli, Senior, Jumper, Doersch, Eslami, Ronneberger, and
  Adler]{rosenbaum2021inferring}
Dan Rosenbaum, Marta Garnelo, Michal Zielinski, Charlie Beattie, Ellen Clancy,
  Andrea Huber, Pushmeet Kohli, Andrew~W. Senior, John Jumper, Carl Doersch,
  S.~M.~Ali Eslami, Olaf Ronneberger, and Jonas Adler.
\newblock Inferring a continuous distribution of atom coordinates from cryo-em
  images using vaes.
\newblock \emph{arXiv preprint arXiv:2106.14108}, 2021.

\bibitem[Shi et~al.(2021)Shi, Luo, Xu, and Tang]{shi2021confgf}
Chence Shi, Shitong Luo, Minkai Xu, and Jian Tang.
\newblock Learning gradient fields for molecular conformation generation.
\newblock In \emph{International Conference on Machine Learning}, 2021.

\bibitem[Xu et~al.(2022)Xu, Yu, Song, Shi, Ermon, and Tang]{xu2022geodiff}
Minkai Xu, Lantao Yu, Yang Song, Chence Shi, Stefano Ermon, and Jian Tang.
\newblock \href{https://openreview.net/forum?id=PzcvxEMzvQC}{{GeoDiff: A
  Geometric Diffusion Model for Molecular Conformation Generation}}.
\newblock In \emph{International Conference on Learning Representations}, 2022.

\bibitem[Jo et~al.(2022)Jo, Lee, and Hwang]{jo2022scorebased}
Jaehyeong Jo, Seul Lee, and Sung~Ju Hwang.
\newblock \href{https://arxiv.org/abs/2202.02514}{{Score-based Generative
  Modeling of Graphs via the System of Stochastic Differential Equations}}.
\newblock \emph{arXiv:2202.02514}, 2022.

\bibitem[Wu et~al.(2022{\natexlab{a}})Wu, Zhang, Jin, Jiang, and
  Li]{wu2022scorebasedmd}
Fang Wu, Qiang Zhang, Xurui Jin, Yinghui Jiang, and Stan~Z. Li.
\newblock A score-based geometric model for molecular dynamics simulations.
\newblock \emph{arXiv preprint arXiv:2204.08672}, 2022{\natexlab{a}}.

\bibitem[Jing et~al.(2022)Jing, Corso, Chang, Barzilay, and
  Jaakkola]{jing2022torsional}
Bowen Jing, Gabriele Corso, Jeffrey Chang, Regina Barzilay, and Tommi Jaakkola.
\newblock Torsional diffusion for molecular conformer generation.
\newblock \emph{arXiv preprint arXiv:2206.01729}, 2022.

\bibitem[Trippe et~al.(2022)Trippe, Yim, Tischer, Baker, Broderick, Barzilay,
  and Jaakkola]{trippe2022motif}
Brian~L. Trippe, Jason Yim, Doug Tischer, David Baker, Tamara Broderick, Regina
  Barzilay, and Tommi Jaakkola.
\newblock Diffusion probabilistic modeling of protein backbones in 3d for the
  motif-scaffolding problem.
\newblock \emph{arXiv preprint arXiv:2206.04119}, 2022.

\bibitem[Hoogeboom et~al.(2022)Hoogeboom, Satorras, Vignac, and
  Welling]{hoogeboom2022equivariant}
Emiel Hoogeboom, Victor~Garcia Satorras, Clément Vignac, and Max Welling.
\newblock Equivariant diffusion for molecule generation in 3d.
\newblock In \emph{International Conference on Machine Learning (ICML)}, 2022.

\bibitem[Anand and Achim(2022)]{anand2022protein}
Namrata Anand and Tudor Achim.
\newblock Protein structure and sequence generation with equivariant denoising
  diffusion probabilistic models.
\newblock \emph{arXiv preprint arXiv:2205.15019}, 2022.

\bibitem[Wu et~al.(2022{\natexlab{b}})Wu, Gong, Liu, Ye, and
  Liu]{wu2022diffusionbased}
Lemeng Wu, Chengyue Gong, Xingchao Liu, Mao Ye, and Qiang Liu.
\newblock Diffusion-based molecule generation with informative prior bridges.
\newblock \emph{arXiv preprint arXiv:2209.00865}, 2022{\natexlab{b}}.

\bibitem[Tuckerman(2010)]{TuckermanBook}
Mark~E. Tuckerman.
\newblock \emph{{Statistical {M}echanics: {T}heory and {M}olecular
  {S}imulation}}.
\newblock Oxford University Press, New York, 2010.

\bibitem[Leimkuhler and Matthews(2015)]{leimkuhler2015molecular}
Benedict Leimkuhler and Charles Matthews.
\newblock
  \emph{\href{https://www.springer.com/gp/book/9783319163741}{{Molecular
  Dynamics: With Deterministic and Stochastic Numerical Methods}}}.
\newblock Interdisciplinary Applied Mathematics. Springer, 2015.

\bibitem[Child(2021)]{child2021very}
Rewon Child.
\newblock Very deep {\{}vae{\}}s generalize autoregressive models and can
  outperform them on images.
\newblock In \emph{International Conference on Learning Representations}, 2021.

\bibitem[Preechakul et~al.(2021)Preechakul, Chatthee, Wizadwongsa, and
  Suwajanakorn]{preechakul2021diffusion}
Konpat Preechakul, Nattanat Chatthee, Suttisak Wizadwongsa, and Supasorn
  Suwajanakorn.
\newblock \href{https://arxiv.org/abs/2111.15640}{{Diffusion Autoencoders:
  Toward a Meaningful and Decodable Representation}}.
\newblock \emph{arXiv:2111.15640}, 2021.

\bibitem[Zhong et~al.(2021{\natexlab{d}})Zhong, Bepler, Berger, and
  Davis]{cryodrgn_zenodo}
Ellen~D. Zhong, Tristan Bepler, Bonnie Berger, and Joseph~H. Davis.
\newblock {Data for "CryoDRGN: Reconstruction of heterogeneous cryo-EM
  structures using neural networks"}, January 2021{\natexlab{d}}.
\newblock URL \url{https://doi.org/10.5281/zenodo.4355284}.

\bibitem[Dormand and Prince(1980)]{dormand1980odes}
J.~R. Dormand and P.~J. Prince.
\newblock
  \href{https://www.sciencedirect.com/science/article/pii/0771050X80900133}{{A
  family of embedded Runge--Kutta formulae}}.
\newblock \emph{Journal of Computational and Applied Mathematics}, 6\penalty0
  (1):\penalty0 19--26, 1980.

\bibitem[Kloeden and Platen(1992)]{kloeden1992euler}
Peter~E. Kloeden and Eckhard Platen.
\newblock \emph{\href{https://www.springer.com/gp/book/9783540540625}{Numerical
  {S}olution of {S}tochastic {D}ifferential {E}quations}}.
\newblock Springer, Berlin, 1992.

\end{thebibliography}

\medskip

\clearpage
\pagebreak

\appendix

\section{Appendix}

\subsection{Implementation and Training Details}\label{app:implementation}
\textbf{CryoDRGN.} All cryoDRGN models are trained as in~\citet{Zhong2021-nmeth}. Dataset and training details summarized in Table~\ref{table:dataset}. Synthetic datasets are generated from atomic models as described in~\citet{zhong2020Reconstructing} and with poses sampled uniformly from $SO(3)$ for rotations and $[-10,10]^2$ pixels for translations. CryoDRGN models for \textit{linear1d} and \textit{circular1d} used positional encodings with geometrically spaced wavelengths between 1 and the Nyquist frequency and leaky ReLU activations. The cryoDRGN model for \textit{ribosome} was downloaded from Zenodo~\cite{cryodrgn_zenodo}. All other settings unless otherwise specified were at their default values for cryoDRGN software version 1.0.

\begin{table}[h]
    \centering
    \caption{Summary of the synthetic and experimental datasets and cryoDRGN training details. $D$ is the resolution of the images, in pixels. $N$ is the dataset size. Architecture is denoted as the width $\times$ depth of the encoder and decoder MLPs.}
    \begin{tabular}{ccccccc}\toprule
    Dataset & $D$ & $N$ & \r{A}/pix. & $|\rvz|$ & Architecture & Epochs  \\
    \midrule
    \textit{Linear1d} & 128 & 50k & 6 & 8 & $256 \times 3$ & 25\\
    \textit{Circular1d} & 128 & 100k & 6 & 8 & $256 \times 3$ & 25 \\
    \textit{Ribosome} (EMPIAR-10076)\cite{Davis2016-ug} & 256 & $87{\small,}328$ & 1.6375 & 10 & $1024 \times 3$ & 50\\
    \textit{Covid} (\citet{Walls2020-yq})& 256 & $276{\small,}549$ & 1.640625 & 8 & $1024 \times 3$ & 25\\
    \bottomrule
    \end{tabular}
    \label{table:dataset}
\end{table}

\textbf{Diffusion Model.} Our latent space diffusion models follow the framework proposed in~\citet{karras2022elucidating}. We set $\sigma_\mathrm{data}$ to the standard deviation of the training dataset. Our model parametrization, including the loss weighting $\lambda(\sigma)$, follows their setup (see column 4 in Table 2 in~\citet{karras2022elucidating}). The neural network backbone of our diffusion models consists of a simple ResNet architecture with 16 hidden layers, using 128 hidden dimensions per layer. We train all models using Adam with a learning rate $3 \cdot 10^{-4}$ and batch size 1024. We set the EMA rate to 0.999.

\textbf{Diffusion Model Sampling.} For deterministic sampling, we use an adaptive step-size Runge--Kutta (RK) 4(5) solver~\citep{dormand1980odes,song2020} for the Probability Flow ODE (\Cref{eq:probability_flow_ode}), with $\sigma(t) = t$. 
To embed points in the latent space of the diffusion model, we run the solver from $\sigma_\mathrm{min}{=}0.002$. to $\sigma_\mathrm{max}{=}80$. On the other hand, for usual sampling (decoding), we run the solver (or sampler) from $\sigma_\mathrm{max}{=}80$ to $\sigma_\mathrm{min}{=}0.002$.

To synthesize regular diffusion model samples, as visualized in~\Cref{fig:prior_samples}, we apply the Euler--Maruyama method~\citep{kloeden1992euler} to the sum of the Probability Flow ODE and a \emph{Langevin diffusion} SDE~\citep{song2020, karras2022elucidating}
\begin{align} \label{eq:diffusion_sde}
    d\rvx = \underbrace{- \dot \sigma(t) \sigma(t) \nabla_\rvx \log p(\rvx; \sigma(t)) \, dt}_{\text{Probability Flow ODE; see~\Cref{eq:probability_flow_ode}}} - \underbrace{\beta(t) \sigma^2(t) \nabla_\rvx \log p(\rvx; \sigma(t)) \, dt + \sqrt{2 \beta(t)} \sigma(t)\, d\omega_t}_{\text{Langevin diffusion component}},
\end{align}
where $d\omega_t$ is the standard Wiener process, $\beta(t)=\dot \sigma(t) / \sigma(t)$, and $\sigma(t)=t$. We use the sampling schedule proposed in~\citet{karras2022elucidating}; in particular:
\begin{align}
    \sigma_i = \left(\sigma_\mathrm{max}^{1/\rho} + \frac{i}{M-1} (\sigma_\mathrm{min}^{1/\rho} - \sigma_\mathrm{max}^{1/\rho}) \right)^\rho, i \in \{0, \dots, M-1\},
\end{align}
with $\rho{=}7.0$. We set the number of function evaluations to M=1,000.

\textbf{Langevin Dynamics.} 
In~\Cref{sec:langevin}, we run regular Langevin dynamics:
\begin{equation}
    \hat{\rvz}_{n+1} = \hat{\rvz}_{n} -\frac{\hat{\rvz}_n}{2}\Delta t + \sqrt{\Delta t}\mathcal{N}(\mathbf{0},\mI),
\end{equation}
using step size $\Delta t = 0.1$. The sequence $\{\hat{\rvz}_n\}$ is then decoded using the RK solver (see above).

\textbf{Latent Interpolations.}
For latent interpolations, we first encode points into the latent space of the diffusion model using the RK solver. We then perform linear interpolation in the diffusion model's latent space. Subsequently, these points are decoded using the RK solver. We generally interpolate using 100 equidistant steps.

\subsection{Additional Plots}
\begin{figure}
    \centering
    \begin{subfigure}[b]{0.49\textwidth}
        \includegraphics[width=\textwidth]{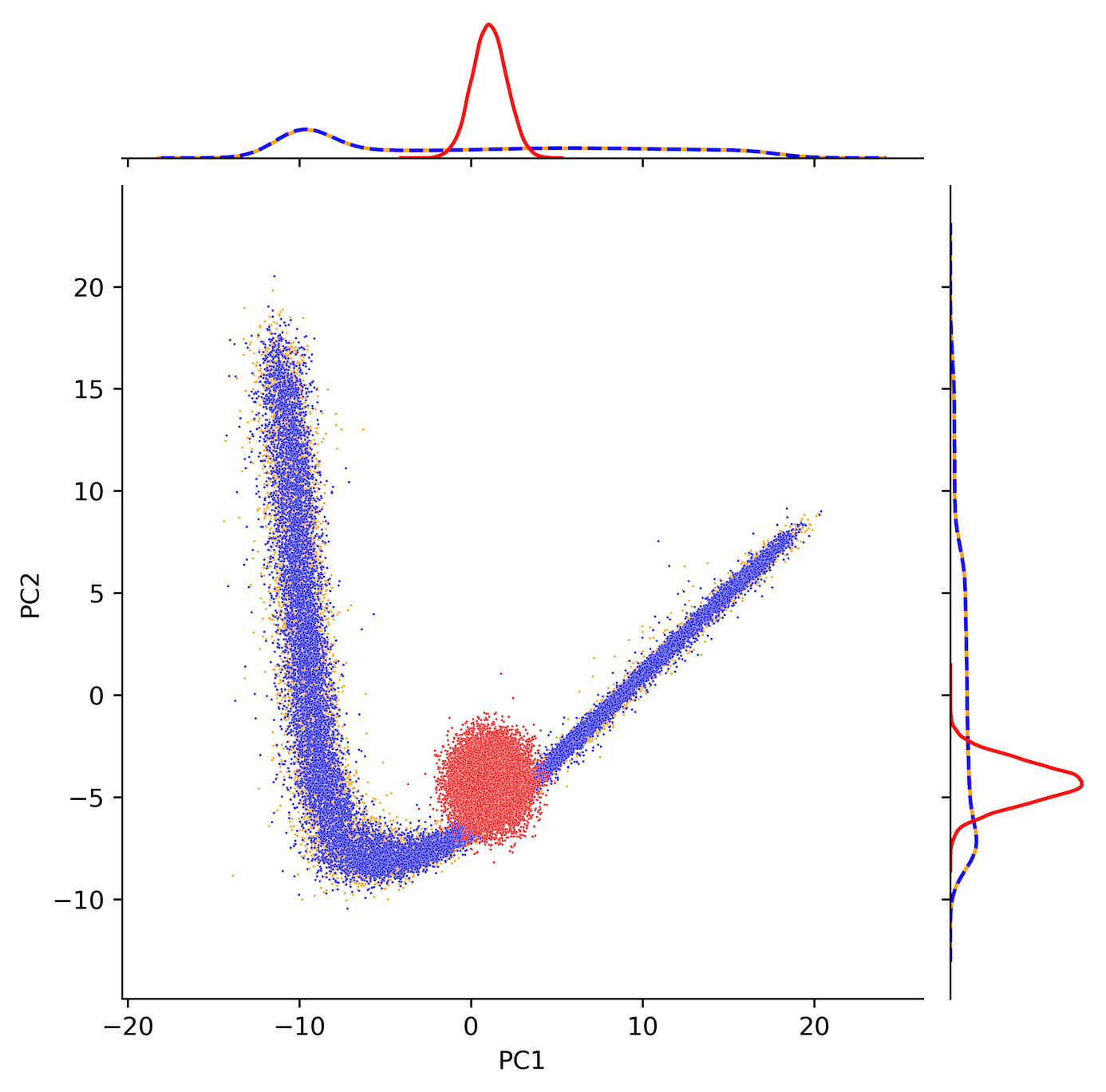}
        \caption{Linear1d}
        \label{fig:marginals_linear}
    \end{subfigure}
    \hfill
    \begin{subfigure}[b]{0.49\textwidth}
        \includegraphics[width=\textwidth]{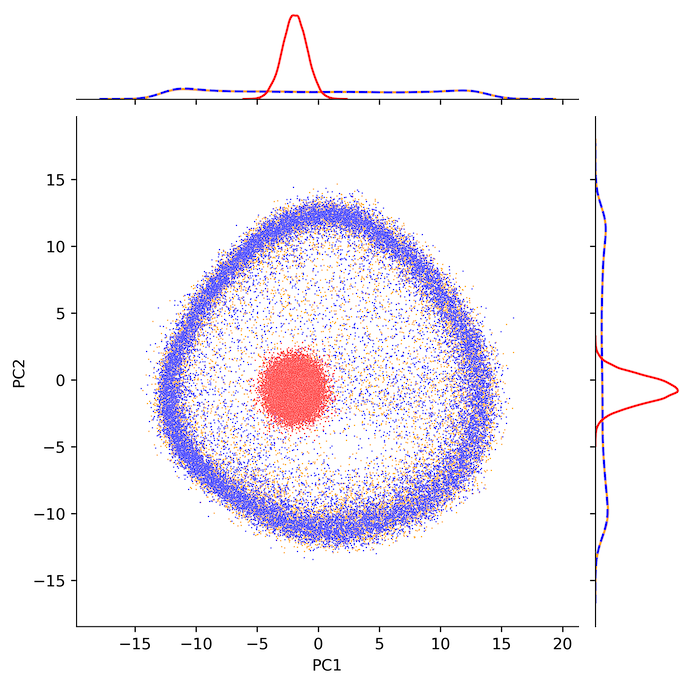}
        \caption{Circular1d}
        \label{fig:marginals_circ}
    \end{subfigure}
    \begin{subfigure}[b]{0.49\textwidth}
        \includegraphics[width=\textwidth]{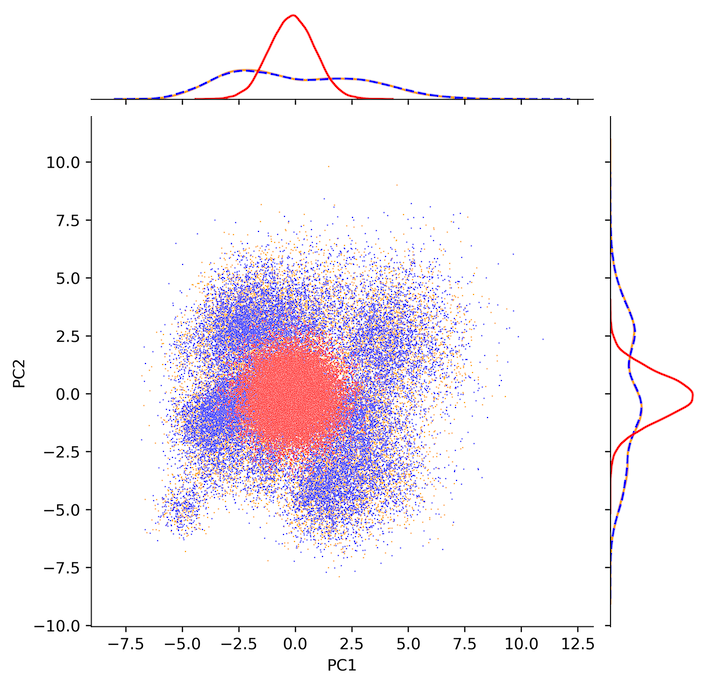}
        \caption{Ribosome}
        \label{fig:marginals_ribo}
    \end{subfigure}
    \hfill
    \begin{subfigure}[b]{0.49\textwidth}
        \includegraphics[width=\textwidth]{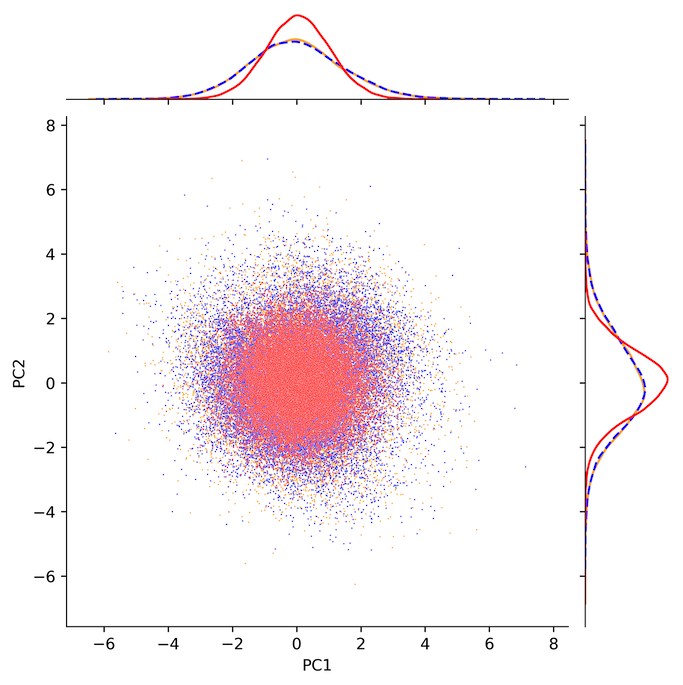}
        \caption{Covid}
        \label{fig:marginals_covid}
    \end{subfigure}
    \caption{Marginal distributions by projecting samples along
the first two principal components of the latent embeddings. Latent embeddings of the data are shown in blue; sampled latent variables from the diffusion model prior or VAE prior
are shown in orange or red, respectively. Note that the data marginals are plotted with a (blue) dashed line, which ends up lying directly on top of the orange curves for the diffusion model samples. From these plots we can observe that the marginal distributions for samples from the diffusion model priors are consistent with those of the data's latent embeddings.}
    \label{fig:marginals}
\end{figure}

In~\Cref{fig:marginals}, we show 1-dimensional marginal distributions of the principal components for all datasets in addition to the samples themselves.

\begin{figure}
    \centering
    \begin{subfigure}[b]{0.49\textwidth}
        \includegraphics[width=\textwidth]{figures/linear_li_colored_dpi200.png}
        \caption{Linear1d}
        \label{fig:latent_interpolation_linear}
    \end{subfigure}
    \hfill
    \begin{subfigure}[b]{0.49\textwidth}
        \includegraphics[width=\textwidth]{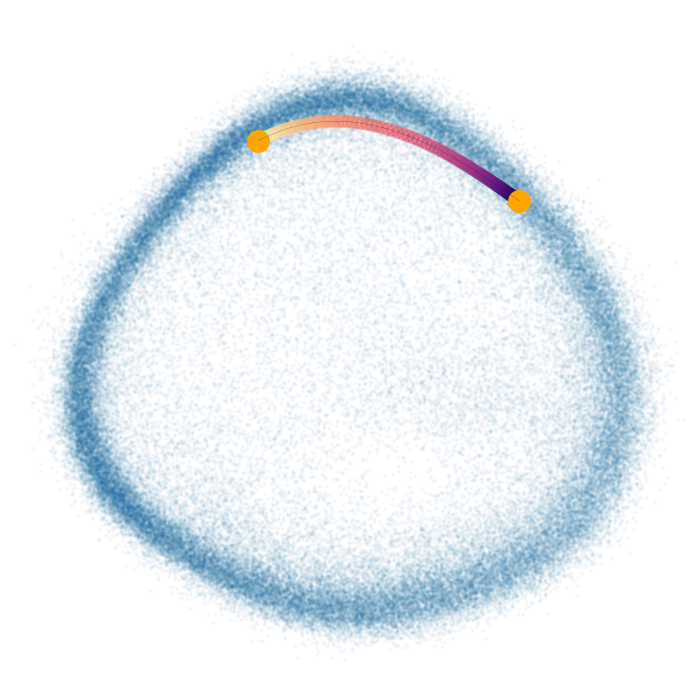}
        \caption{Circular1d}
        \label{fig:latent_interpolation_circ_v2}
    \end{subfigure}
    \caption{Latent interpolation in the diffusion model prior visualized using PCA (as in \cref{fig:marginals}); data latent embeddings are shown in blue. Starting and end points are shown in orange. Zoom in for details.}
    \label{fig:latent_interpolation}
\end{figure}

In~\Cref{fig:latent_interpolation}, we show a conformational state interpolation for \textit{circular1d}. Leveraging the diffusion model prior, we see that the interpolation proceeds along the data manifold.

\begin{figure}
    \centering
    \begin{subfigure}[b]{0.49\textwidth}
        \includegraphics[width=\textwidth]{figures/linear_ld_colored_dpi150.png}
        \caption{Linear1d}
        \label{fig:langevin_dynamics_linear}
    \end{subfigure}
    \hfill
    \begin{subfigure}[b]{0.49\textwidth}
        \includegraphics[width=\textwidth]{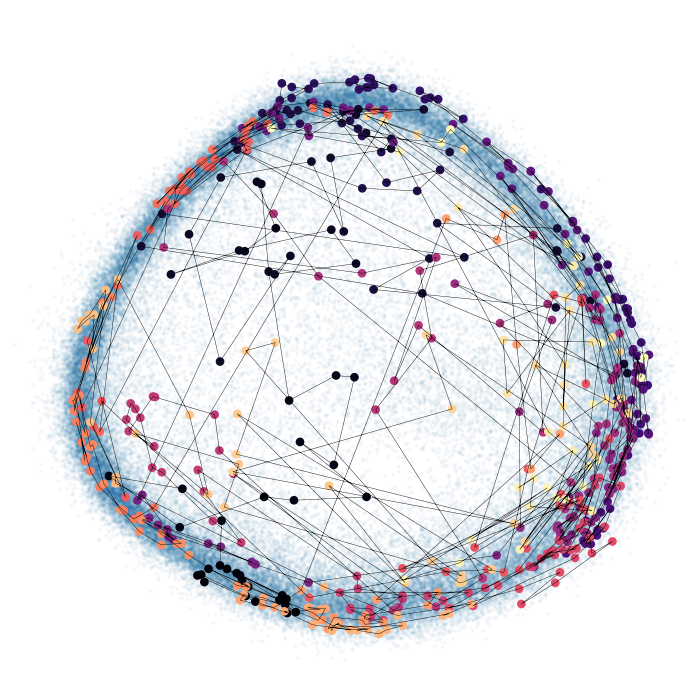}
        \caption{Circular1d}
        \label{fig:langevin_dynamics_circ}
    \end{subfigure}
    \begin{subfigure}[b]{0.49\textwidth}
        \includegraphics[width=\textwidth]{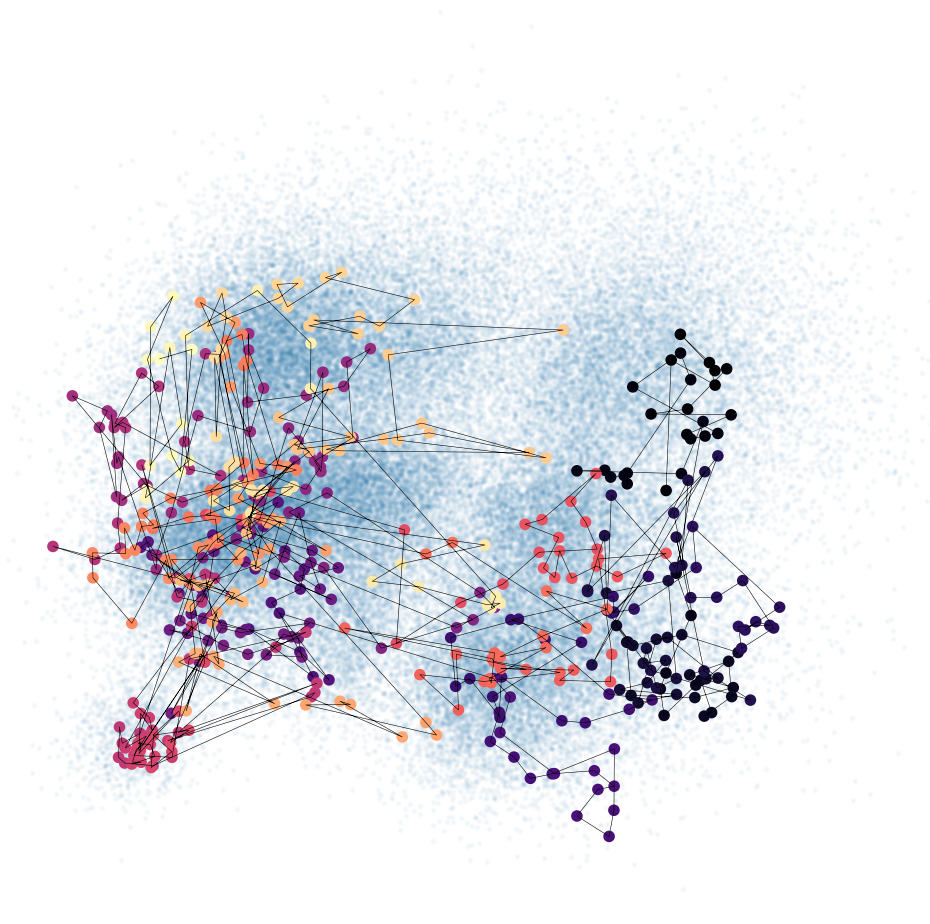}
        \caption{Ribosome}
        \label{fig:langevin_dynamics_ribo}
    \end{subfigure}
    \hfill
    \begin{subfigure}[b]{0.49\textwidth}
        \includegraphics[width=\textwidth]{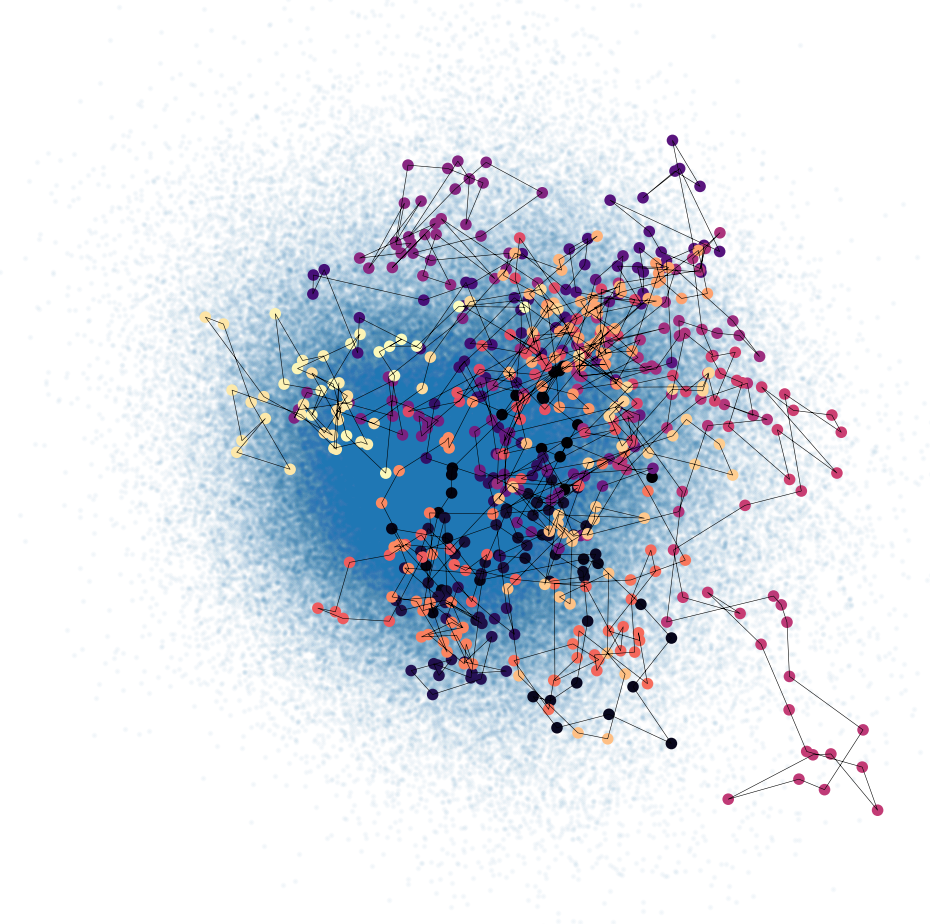}
        \caption{Covid}
        \label{fig:langevin_dynamics_covid}
    \end{subfigure}
    \caption{Langevin dynamics in the diffusion model prior visualized using PCA (as in \Cref{fig:marginals}). Data latent embeddings are shown in blue. The color-coding indicates the sampling path---color changes smoothly as the Langevin dynamics proceed. Zoom in for details.}
    \label{fig:langevin_dynamics}
\end{figure}

In~\Cref{fig:langevin_dynamics}, we show fast Langevin dynamics sampling leveraging the diffusion model prior for all four datasets.


\end{document}